\documentclass[prb,twocolumn,showpacs,amssymb,amsmath,floatfix]{revtex4}
\usepackage{epsfig}
\usepackage{graphicx}
\usepackage{xcolor}

\begin{document}

\title{
Dependence of transport coefficients of Yb(Rh$_{1-x}$Co$_x$)$_2$Si$_2$ intermetallics  
on temperature and cobalt concentration 
}

\author{V.~Zlati\'c}
\affiliation{Department of Physics, Faculty of Science, University of Split, R. Bo\v skovi\'ca 33, 21000 Split, Croatia}
\author{U.~Stockert}
\affiliation{MPI for Chemical Physics of Solids, D-01187 Dresden, Germany}

\begin{abstract}
Dependence of transport coefficients of the Yb(Rh$_{1-x}$Co$_x$)$_2$Si$_2$ series of alloys on temperature 
and cobalt concentration is explained by an asymmetric Anderson model which takes into account the exchange scattering 
of conduction electrons on ytterbium ions and the splitting of 4$f$-states by the crystalline electric field (CEF). 
The substitution of rhodium  by cobalt is described as an increase of chemical pressure 
which reduces the exchange coupling and the CEF splitting. 
The scaling analysis and numerical NCA solution  of the model show that the effective degeneracy 
of the 4$f$-state at a given temperature depends on the relative magnitude of the Kondo scale and the CEF 
splitting. 
Thus, we find that dependence of the thermopower, $S(T)$, on  temperature and cobalt concentration 
can be understood as an interplay of quantum fluctuations, driven by the Kondo effect, 
and thermal fluctuations, which favor a uniform occupation of the CEF states.  
The theoretical model captures all the qualitative features of the experimental data and 
it explains the evolution of the shape of $S(T)$ with the increase of cobalt concentration. 

\end{abstract}


\maketitle



\section{Introduction
\label{introduction}  
}
Recent investigations of electrical, thermal, magnetic, and spectroscopic properties of the strongly correlated 
material Yb(Rh$_{1-x}$Co$_x$)$_2$Si$_2$, for various cobalt concentrations, revealed a number of interesting 
features that have not been completely 
understood\cite{stockert.2019,Klingner.2011,Kummer.2018,Goremychkin1982,Stockert2006}. 
The functional form of transport coefficients changes very much as  rhodium is replaced by cobalt: 
The thermopower exhibits a single broad minimum at the rhodium-rich end 
and it acquires an additional minimum for large enough cobalt concentration\cite{stockert.2019}. 
The shape of the resistivity also changes as cobalt concentration increases\cite{Klingner.2011}.
The transport coefficients of several other ytterbium-based intermetallic compounds 
show similar evolution, when substituting one element by another or applying 
hydrostatic pressure\cite{TB-99-3,YRS-08-4,TB-12-2,TB-99-4}. 
Some of the observed features are typical of Kondo systems but the overall temperature dependence 
of the data cannot be characterized by a single temperature scale\cite{Klingner.2011,stockert.2019}. 
Here, we show that the thermopower of the Yb(Rh$_{1-x}$Co$_x$)$_2$Si$_2$ series of alloys 
can be explained by the Kondo  scattering of conduction electrons on the rare earth ions but 
that we also have to include the effects of the crystalline electrical field (CEF). 
The Kondo scale depends on the effective degeneracy of the 4$f$ state which can 
be different at low and high temperatures, because of the CEF splitting\cite{zlatic.2014}, 
while the temperature at which the degeneracy changes depends on cobalt concentration. 
An interplay of the Kondo effect and the CEF splitting can also be seen in the transport data 
of cerium compounds taken at various pressure or chemical pressure.\cite{wilhelm.2005,ocko.2001} 

The resonant x-ray emission spectroscopy of the Yb(Rh$_{1-x}$Co$_x$)$_2$Si$_2$ series 
of alloys indicates\cite{Kummer.2018} that Yb ions fluctuate between the ${2+}$ and ${3+}$ 
configurations, as one would expect of a Kondo system. The spectroscopic weight of  Yb$^{2+}$ 
decreases gradually from the rhodium-rich to cobalt rich end of the series. This is consistent with 
the transport\cite{Klingner.2011,stockert.2019} and thermodynamic\cite{Klingner.2011} data which 
show that the low-temperature Kondo scale drops from about 25 K in YbRh$_2$Si$_2$ to less than 1 K 
in YbCo$_2$Si$_2$. 
The reduction of the Kondo scale with Co concentration, i.e., an increase of the magnetic character 
of Yb ions, indicates that an increase of cobalt gives rise to chemical pressure which stabilizes the Yb$^{3+}$ configuration.
The overall temperature dependence of the number of 4$f$ holes inferred 
from the spectroscopic data appears to be nonuniversal\cite{Kummer.2018}.
(By the "nonuniversality", we mean that the data cannot be characterised by a single temperature scale.)

The neutron scattering data\cite{Stockert2006} reveal that the 4$f$-state of YbRh$_2$Si$_2$  
is split by the CEF into four doublets and that the first excited level is separated from 
the ground state by an energy $\Delta_1\simeq 200$ K.  
In YbCo$_2$Si$_2$, one finds the same level scheme\cite{Goremychkin1982} but 
with reduced overall CEF splitting and $\Delta_1\leq $50 K. 
Magnetic and thermodynamic measurements on Yb(Rh$_{1-x}$Co$_x$)$_2$Si$_2$ alloys 
provide indirect evidence\cite{Klingner.2011,Kummer.2018} that $\Delta_1$ decreases as $x$ increases but 
that the structure of the CEF excitations doesn't change throughout the series.
Thus, an increase of cobalt concentration reduces not just the Kondo coupling 
but the CEF splitting as well; this should be taken into account, when analyzing the thermoelectric 
response of these materials. 

This paper explains the thermopower of the Yb(Rh$_{1-x}$Co$_x$)$_2$Si$_2$  series 
of alloys\cite{stockert.2019} using a highly asymmetric Anderson model with several $f$-like multiplets. 
The calculations take into account the reduction of the exchange coupling and the CEF splitting 
due to the substitution of rhodium by cobalt. 
The model neglects all the details of the band structure, so we only aim at a qualitative description 
of the experimental results. The fact that the Nordheim-Gorter rule doesn't hold and that the single 
ion contribution is difficult to extract makes a quantitative comparison between 
theory and experiment rather difficult. 

Our theoretical results capture all the qualitative features 
of the experimental data on Yb(Rh$_{1-x}$Co$_x$)$_2$Si$_2$. In what follows, 
we discuss the experimental results in Sec. \ref{experiment}, describe the model in Sec. \ref{model}, 
provide a qualitative solution obtained by a poor man's scaling  in Sec. \ref{pms},  
and discuss the numerical NCA solution of the model in Sec. \ref{NCA}.    
The theoretical and experimental results are compared in Sec. \ref{comparioson}. 
The summary is provided in Sec. \ref{conclusions}. 

\section{Experiment
\label{experiment}  
}
\begin{figure}
\center{\includegraphics[width=.9 \columnwidth,clip] {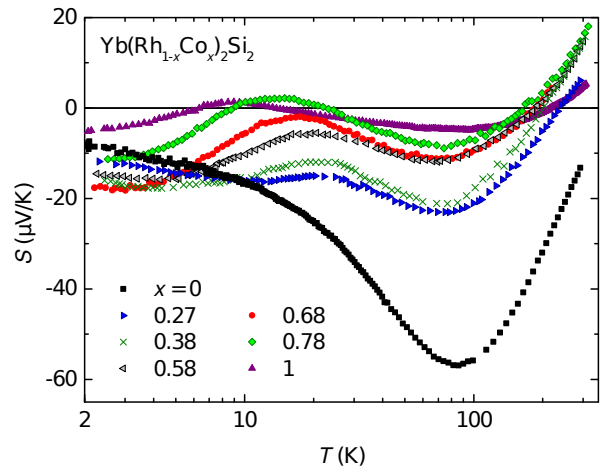}}
\caption{Thermopower of Yb(Rh$_{1-x}$Co$_x$)$_2$Si$_2$ plotted as a function of temperature 
for various values of cobalt concentration.
 }
                                  \label{fig:thermopower}      
\end{figure}

The temperature dependence of the thermopower of Yb(Rh$_{1-x}$Co$_x$)$_2$Si$_2$ alloys,  
 $S(T)$, is shown in Fig.\ref{fig:thermopower} for various concentrations of cobalt ions. 
At low cobalt concentration,  $x < 0.195$, the thermopower exhibits a deep minimum at temperature $T_\mathrm{min}\simeq 100$ 
where $S(T_\mathrm{min})=-60 ~ \mu V/K$. Above $ T_\mathrm{min}$, the thermopower increases logarithmically 
and, for $x>0$, it changes sign around room temperature.  
An increase of $x$ reduces the depth of the minimum but doesn't change the value of $T_\mathrm{min}$ much. 
A thermopower of a similar shape is often found in Kondo systems with a large Kondo scale or 
in valence fluctuating materials.\cite{zlatic.05}  

At higher cobalt concentration, $x \geq 0.27$, the thermopower  acquires a double-well shape, i.e., 
in addition to the high-temperature minimum, around $T_\mathrm{min}\simeq$100 K, 
it develops a low-temperature one at $T_0\leq 10$ K. 
For temperatures between $T_{0}$ and $T_\mathrm{min}$, we find a maximum of $S(T)$ at temperature $T_\mathrm{max}$, 
where $S(T_\mathrm{max})$ is negative for $x\leq 0.68$ and positive for $x>0.68$.  
An increase of $x$ pushes $T_\mathrm{max}$ and $T_0$ to lower values; 
it also reduces temperature at which $S(T)$ changes sign. 
Contrary to that,  $T_\mathrm{min}$ doesn't change much with cobalt concentration. 
 
The electrical resistivity of Yb(Rh$_{1-x}$Co$_x$)$_2$Si$_2$ alloys\cite{Klingner.2011}  exhibits 
a maximum at temperature $T_\mathrm{max}^{\rho}$ which decreases as the concentration of cobalt increases.  
For large enough $x$, a secondary maximum appears at a higher temperature 
which is not very different from $T_\mathrm{min}$. 
For $T\geq T_\mathrm{max}^{\rho}$, the conduction electrons' mean free path is very short 
(comparable to the lattice spacing), so that the transport properties can be discussed  
assuming incoherent scattering of conduction electrons on Yb ions. 
Since we always find $T_\mathrm{max}^{\rho} <T_0$, the thermopower of Yb(Rh$_{1-x}$Co$_x$)$_2$Si$_2$ 
can be described by an impurity model at any $x$.  
For a periodic array of Yb ions, such a description breaks down at $T \ll T_\mathrm{max}^{\rho}$ due to 
the onset of coherence. At such low temperatures, the resistivity and thermopower 
should follow the Fermi liquid (FL) laws.\cite{Hewson}

\section{The model
\label{model}  }

The model of Yb(Rh$_{1-x}$Co$_x$)$_2$Si$_2$ alloys considers one Yb ion per unit cell 
and assumes that the scattering of conduction holes on a given Yb ion is independent of 
other Ytterbiums, except through the modification of the chemical potential.
In other words, we use an effective impurity model which treats the 4{\it f} states as
scattering resonances rather than Bloch states but it allows for the charge transfer 
between the conduction and $f$-states\cite{Bickers1987,Hewson,costi.94,TB-97-1,zlatic.05,zlatic.2014}. 
The chemical potential, $\mu$, is adjusted at every temperature and for every value of $x$, 
so as to maintain the overall charge neutrality of the compound.
That is, for $n_c$ conduction holes and $n_f$  Yb holes per site, the total charge $n=n_c+n_f$ 
is always conserved. 
The fluctuations of Yb ions between the low-energy 4$f^{13}$ and the high-energy 4$f^{14}$ or 4$f^{12}$ 
configuration broaden the 4{\it f}-states into resonances of half-width $\Gamma=\pi V^2 N(\mu)$, where $V$ 
is the hybridization matrix element and $N(\mu)$ is the conduction electrons density of states at the Fermi level. 

In the large-$U$ limit, where the two-hole configuration of Yb ions can be neglected, 
we arrive at  an effective single-impurity Anderson model in which the single-hole configuration is 
represented by ${N}-1$ excited CEF states separated from the ground state by energies $\Delta_i=E_f^{i}-E_f^{0}$ 
($i=1,\ldots,{ N} -1$).
Here,  $E_f^{0}$ and $E_f^{i}$ are the energies of the ground and excited states, respectively. 
The total degeneracy of 4$f$-states is $ {\cal N}=\sum_{i=0}^{{ N} -1} {\cal N}_i $, where 
the individual  degeneracies of the CEF states, ${\cal N}_i$, are determined  by the point group symmetry 
of the crystal. The Kondo effect due to the 4$f^{13}$ -- 4$f^{14}$  fluctuations 
is described by the  Hamiltonian,\cite{Bickers1987}
\begin{equation}
                               \label{Hamiltonian}
H_{A}=H_\mathrm{band}+H_\mathrm{imp}+H_\mathrm{mix}~.
\end{equation}
$H_\mathrm{band}$ describes the single-hole excitations in the conduction band with a semielliptical density of states  
centered at $E_c^0$ and of half-width $W$, which we take as the unit of energy. 
The origin of the energy axis is set to $\mu$. 
The unrenormalized CEF excitations of 4{\it f} ions are described  by $H_\mathrm{imp}$; 
their spectral functions are given by a set of delta functions at energies $E_f^{i} > 0$ ($i=0,\ldots,{ N}$). 
The transfer of holes between the 4{\it f} ions and conduction band is described by $H_\mathrm{mix}$ 
which depends on the coupling constant $g_0=\Gamma/\pi E_f^0$. 
This coupling broadens the spectral functions of 4$f$ excitations into narrow resonances. 

The qualitative features of the model depend in an essential way on the coupling constant $g_0$ 
and the CEF splitting $\Delta_1$.  In Yb intermetallics, pressure or chemical pressure shift the 
4$f$-states away from the chemical potential, without changing their width $\Gamma$ much. 
Thus, an increase of cobalt concentration increases $E_f^0$, which reduces the coupling constant 
and the Kondo scale.  
The data on Yb(Rh$_{1-x}$Co$_x$)$_2$Si$_2$ alloys also show that an increase of cobalt concentration 
reduces the CEF splitting. A smaller splitting facilitates the thermal occupation of excited states and affects 
the degeneracy of the 4$f$-states at a given temperature.

In what follows, we first analize the model by the poor man's scaling and obtain its qualitative features  
in various parts of the parameter space. Then, we find the solution by the non crossing approximation (NCA) 
and study in detail the effects of temperature and cobalt concentration on the thermopower. 
The input parameters for numerical calculations are $\Gamma$, $E_f^0$ ($g_0=\Gamma/\pi E_f^0$), 
$\Delta_i$,  ${\cal N}_i$, and the total number of electrons $n_f+n_c$. For Yb-based compounds, 
we have $n_f\leq 1$ and assume $\Gamma, \Delta_1 \ll E_f^0  < W$, i.e., $g_0\ll 1$. 

\section{The scaling solution \label{pms}  }
For given  input parameters, the scaling equations are obtained by reducing the band 
width from $W$ to $W-\delta W$ and renormalizing the coupling constant from $g_0$ to $g_0+\delta g$, 
while keeping the form of the Hamiltonian unchanged. 
Considering, for simplicity, a $m$-fold degenerate CEF ground state separated from a $M$-fold 
degenerate excited state by energy $\Delta$, we obtain the scaling equation
\cite{hanzawa.85,zlatic.05},  
\begin{equation}
                               \label{scaling}
\left({T_K\over W}\right)^{m/{\cal N}} \left( {T_K+\Delta} \over W \right)^{M/{\cal N}} 
= \left( {T_K^0\over W} \right)^{2/{\cal N}} ~, 
\end{equation} 
where $T_K$ is the scaling invariant (Kondo scale) of the model at hand, ${\cal N}=m+M$, 
and $T_K^0 = W e^{-1/2 g_0}$ is the scaling invariant of a two-fold degenerate model with 
the same bare coupling $g_0$ and unrenormalized band-width $W_0$.
The Kondo scale is defined as the low energy cut-off at which the renormalized coupling 
diverges\cite{Anderson.70}. The scaling equation shows that $T_K$ depends  exponentially 
on the coupling constant and algebraically on the CEF splitting. 

In the scaling regime, the properties of the system are determined by the renormalized coupling, 
$g(W,T_K,\Delta)$, which is obtained by replacing $g_0$ by $g$ in Eq.\eqref{scaling}. 
Then, for a given $T_K$ and $\Delta$,  we define the scaling trajectory $g(T)$ by making 
the replacement $W\to T$. 
Along the scaling trajectory, the transport coefficients are obtained from the lowest order perturbation 
theory\cite{Bickers1987,zlatic.2014}  in which the bare exchange coupling $g_0$ is replaced by the 
renormalized one $g(T)$.    
The results obtained in such a way are equivalent to what one finds in an infinite-order perturbation theory 
which uses the unrenormalized coupling constant and keeps the parquet diagrams\cite{Hewson,zlatic.2014}. 
Depending on the relative magnitude of $T_K$ and $\Delta$, the transport coefficients 
generated by scaling theory exhibit qualitatively different features,  

If the initial parameters are such that  $T_K$ is larger or comparable to $T_\Delta$, 
where $ T_\Delta = \Delta/A$ and $A\simeq $ 3 -- 5, 
the quantum fluctuations render the CEF splitting irrelevant 
and the system behaves as a $\cal N$-fold degenerate one. 
In that case, the approximate solution of Eq.\eqref{scaling} is given  by  
\begin{equation}
                               \label{}
{T_K^{\cal N}\over W}= \left[{T_K^0\over W}\right]^{2\over {\cal N} } ~ , 
\end{equation} 
where $T_K^{\cal N}$ is the Kondo scale of an ${\cal N}$-fold degenerate $f$-state 
coupled to a conduction band of halfwidth $W$. 
Since ${\cal N}$ is large, $ T_K^{\cal N}$ is greatly enhanced with respect to $T_K^0$. 

At high temperatures, $T/T_K^{\cal N}\geq 1$, the system with a large coupling constant is in 
the local moment (LM) regime, where we can calculate the correlation functions by the renormalized perturbation theory. 
The lowest order expansion in terms of $g(T)$ gives for the electrical resistivity a universal function 
of reduced temperature $t=T/T_K^{\cal N}$.  
The resistivity $\rho=\rho\left[ g(t) \right]$ is a monotonic function of $t$: 
It is negligibly small for $t\gg 1$ and it is close to the unitarity limit for $t\simeq 1$.  
(The unitarity limit of the Anderson model is defined by $\rho(t=0)$ and it is known exactly 
from the phase shift analysis and the Friedel sum rule.\cite{Bickers1987,Hewson}) 
The high-temperature thermopower of Yb compounds, obtained in the same way, has a positive slope.  
At low temperatures, $t\leq 1$, the renormalized perturbation expansion cannot be used, 
because $g(T)$ becomes bigger than one, but the exact calculations\cite{Hewson}  show that the 
system undergoes, for $t\simeq 1$, a crossover to a non-degenerate FL state\cite{hanzawa.85,zlatic.2014}. 
Since the thermopower has to vanish at $T=0$, its low-temperature slope is negative.
Thus, in the case of an Yb compound with large exchange coupling,  $T_K^{\cal N}\geq T_\Delta$,  
we expect the thermopower with a single (negative) minimum at temperature $T_\mathrm{min}$ 
which indicates the LM--FL crossover. 
A decrease of the coupling constant, for given $\cal N$, reduces $T_K^{\cal N}$  
and shifts $T_\mathrm{min}$ to lower temperatures. 

In the opposite case, such that $T_K\ll \Delta$, the approximate solution of Eq.\eqref{scaling} 
can be written as 
\begin{equation}
                               \label{}
{T_K\over \Delta }=   \left({T_K^{\cal N} \over \Delta} \right)^{N/m}~ , 
\end{equation} 
or
\begin{equation}
                               \label{}
{T_K\over W}= \left(W\over {\Delta}\right)^{M/ 2} ~{T_K^0\over W}~.
\end{equation}  
Since $T_K^{\cal N}/\Delta\ll 1$ and $W/\Delta\gg 1$, we have $T_K^0\ll T_K\ll T_K^{\cal N}\ll \Delta$. 
Here, $T_K$ is the Kondo scale of an impurity with $M$-fold degenerate excited states 
split-off from the $m$-fold degenerate ground state by energy $\Delta$. 
The value of $T_K$ is enhanced with respect to the Kondo scale of a plain doublet, $T_K^0$,  
by factor $ \left(W/{\Delta}\right)^{M/ 2}$. 
An increase of $g$ enhances $T_K$ exponentially fast, while an increase of $\Delta$ 
reduces $T_K$ as a power law.  Thus, for large enough coupling, the $f$-state 
behaves as a $\cal N$-fold degenerate one, i.e., the quantum fluctuations render 
the CEF splitting irrelevant. 

At high temperatures,  $T > T_\Delta$, thermal fluctuations give rise to a uniform occupation of all the 
CEF states and the system is effectively $\cal N$-fold degenerate, with the scaling invariant $ T_K^{\cal N}$. 
For $ T\leq T_\Delta$, the high-temperature scaling terminates when the excited CEF states depopulate 
and  the system makes a crossover to a low-temperature LM regime. 
For $T \ll T_\Delta$, the {\it f}-state behaves as an effective doublet (or quartet) with Kondo scale $T_K$, 
which is enhanced with respect to $T_K^0$ by quantum fluctuations, 
i.e., by virtual transitions from the ground state doublet into the unoccupied CEF states. 
For $T < T_K$, the remaining magnetic entropy of the twofold (or $m$-fold) degenerate LM 
is removed by the crossover into the FL state. Thus, the system is characterized by two 
crossovers: The high-temperture one, between the two LM regimes, and the low-temperature one 
into the FL ground state. 

The transport coefficients in each of the scaling regimes can be inferred 
from the renormalized perturbation theory. 
For $T\geq T_\Delta$,  the resistivity and the thermopower are universal functions  
of $T/T_K^{\cal N}$, while for $T \ll T_\Delta$, they are universal functions  of $T/T_K$. 
Since $T/T_K^{\cal N}\ll T/T_K$ and the resistivity monotonically decreases as temperature increases, 
we have $\rho(T/T_K) <\rho(T/T_K^{\cal N})$, i.e.,  the resistivity drops at the crossover 
from large to small values. 
Below the crossover, the resistivity increases again as temperature decreases, albeit with a  smaller slope. 
This increase is due to the Kondo scattering of conduction electrons on the effective 4{\it f} doublet. 
Eventually, at low enough temperatures, rather than saturating at the unitarity limit for impurity scattering, 
the resistivity of intermetallic compounds drops to very small residual values characteristic of a lattice 
of coherent 4$f$ states. 
Thus, in the case $T_K\ll T_K^{\cal N}\ll \Delta$, the resistivity exhibits two well resolved maxima. 

With regard to the thermopower of a system with $T_K\ll T_K^{\cal N}\ll \Delta$, 
it can assume large values at high temperatures, $T >\Delta$, where a single 4{\it f}~hole 
is distributed over $\cal N$ scattering channels and the system is far from electron-hole symmetry. 
Since the thermopower is a universal function of $T/T_K^{\cal N}$ and we have $T\gg T_K^{\cal N}$, 
the slope of $S(T)$ in this high-temperature LM regime is positive. 
The reduction of temperature can bring $S(T)$ down to rather large negative values, 
typical of Kondo systems with large Kondo scale. 
Eventually, for $T \leq T_\Delta$, thermal depopulation of excited CEF states gives rise 
to the crossover into the low-temperature LM regime, where the thermopower is small, 
because a singe $f$-hole has only two channels available for scattering and the system 
is close to half-filling. (At half filling exactly, $S(T)=0$.)  
In the crossover regime $S(T)$ has a negative slope, such that the break-down 
of the $\cal N$-fold degenerate LM regime is characterized by a minimum of $S(T)$ 
at $T_\mathrm{min} < T_\Delta$. Note,  $T_\mathrm{min}$ is not related to $T_K^{\cal N}$ but it 
indicates the onset of the crossover between two LM regimes which occurs because 
of thermal depopulation of the excited CEF states.  
 
Once the low-temperature LM regime is stabilized, i.e. temperature is such that $T_K\ll T \ll T_\mathrm{min}$, 
the thermopower becomes a universal function of $T/T_K$ with a positive slope. 
Eventually, for $T\simeq T_K$, the magnetic entropy of the m-fold degenerate state 
is removed by a crossover into the FL regime, where $S(T)\simeq - T/T_K$. 
Thus, the poor man's scaling of the Anderson model with the CEF splitting yields 
the thermopower with two minima: the high-temperature one, indicating the collapse 
of the $\cal N$-fold degenerate LM regime and the low-temperature one, indicating the 
collapse of the $m$-fold degenerate LM regime, and the crossover into the nondegenerate FL regime. 

The scaling theory describes rather accurately the properties of the system in the LM regimes. 
However, it cannot describe  the crossovers or the low-temperature FL regime. 
Under chemical pressure, the coupling constant and the CEF splitting of  Yb(Rh$_{1-x}$Co$_x$)$_2$Si$_2$
change simultaneously, making the scaling trajectory subject to two competing effects. 
The ensuing shifts of the thermopower minima are difficult to infer from scaling and 
in order to analyze the experimental data, a more accurate solution of the Anderson model 
has to be used. This is provided by the NCA calculations which we present next. 

\section{The numerical NCA solution
\label{NCA}  }

The single-particle spectral functions of the {\it f}-holes and the transport coefficients of the 
Anderson model inferred from the numerical solution of the  NCA equations have been discussed in 
Refs.[\onlinecite{Bickers1987,Hewson,zlatic.05}].  The effect of chemical pressure on 
Yb(Rh$_{1-x}$Co$_x$)$_2$Si$_2$ is taken into account by simultaneously changing 
the coupling constant and the CEF splitting at each cobalt concentration. We consider a model 
with a ground state doublet at energy $E_f^0$, an excited quartet at $E_f^0+\Delta_1$, and an 
additional excited doublet at $E_f ^0+\Delta_2$, where all the energies are measured 
with respect to $\mu$.\footnote[1]{We consider a model with an excited quartet rather than 
two excited doublets to facilitate the numerical work.}
 
To illustrate the behavior of the spectral function we choose $E_f^0=1.0$ eV, $\Delta_1=0.06$ eV, 
and $\Delta_2=1.06$ eV. 
The dependence of thermopower on temperature and cobalt concentration is obtained by performing 
the calculations  for several pairs of $E_f^0$ and $\Delta_1$. The 4$f$ resonance and conduction 
band have the half-width $\Gamma=0.12$ eV and $W$=4 eV, respectively. We take $n_\mathrm{tot}$=5.63 
electrons per ion, i.e., there are (on average) 0.74 holes in each one of eight  hybridized channels.

\begin{figure}
\center{\includegraphics[width=.8 \columnwidth,clip] {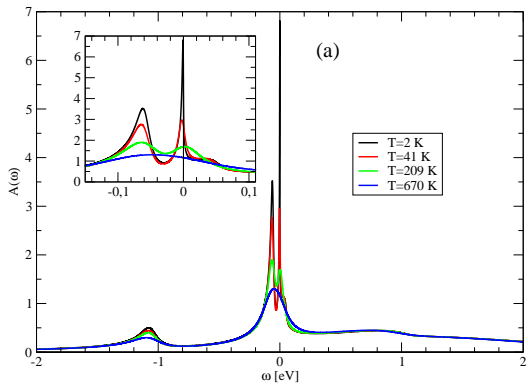}}
\caption{$f$-electron spectral function $A(\omega)$, calculated by the NCA, plotted as a function of 
frequency for several  temperatures, as indicated in the figure. 
The data show the results for an Yb ground state doublet at $E_f^0=1.0$ eV,  
an excited quartet at $E_f^0+\Delta_1$, and a doublet at $E_f ^0+\Delta_2$, where $\Delta_1=0.06$ eV and $\Delta_2=1.06$ eV. 
The hybridisation strength $\Gamma=0.12$ eV  and the conduction band halfwidth is 4 eV.  
Inset: The low-frequency part of the spectral functions. 
  }
                                  \label{fig:spectrum_Yb_CF_case}      
\end{figure}
\subsection{Spectral function\label{spectrum}  }
The spectral function, $A(\omega)$, of an Yb 4$f$-hole  is shown in Fig.\ref{fig:spectrum_Yb_CF_case} 
as a function of frequency, for several temperatures. 
At low temperatures, it exhibits a Kondo resonance centered slightly below the chemical potential. 
In the NCA calculations, the Kondo scale of the model is provided by the distance of the center of the 
resonance from the chemical potential\cite{Bickers1987,Hewson}. 
For the model with the CEF splitting, this distance is exponentially reduced with respect 
to the Kondo scale of an eightfold degenerate model with the same coupling constant. 
The spectral function exhibits additional low-energy resonances which are shifted by $\pm \Delta_1$  
from the chemical potential. These resonances are due to the exchange scattering of conduction electrons 
on the excited CEF quartet. The weak resonance at about $\omega\simeq 0.95$ eV is due to the 
single particle excitations and its position is determined by $E_f^0$. (The small downward shift with 
respect to the input value is caused by enforcing the charge neutrality in each unit cell.)  
The  weak resonance at $\omega\simeq -1.1$ eV is also a many-body effect due to the highest CEF doublet.

The temperature dependence of the low-energy spectral weight is shown in the inset 
of Fig. \ref{fig:spectrum_Yb_CF_case}. 
For $T\leq \Delta_1/3$, the  peaks due to the CEF excitations are well resolved but 
for $T > \Delta_1$ only a single Kondo resonance remains (blue curve). 
It characterizes an eightfold degenerate $f$-hole and if the distance of the resonance from the chemical 
potential is used to estimate the high-temperature Kondo scale, $T_K^{\cal N}$, the values thus obtained 
are consistent with the NCA results for an octet. 
The temperature variation of $A(\omega)$ clearly indicates two crossovers. The first one, at high 
temperatures, is from an eight-fold degenerate LM regime with large Kondo scale to the two-fold degenerate 
LM regime with small Kondo scale. The second one, at low-temperatures, is from the 
 two-fold degenerate LM regime to the non-degenerate FL ground state. 

\subsection{Thermopower\label{thermopower}  }
\begin{figure}
\center{\includegraphics[width=.8 \columnwidth,clip] {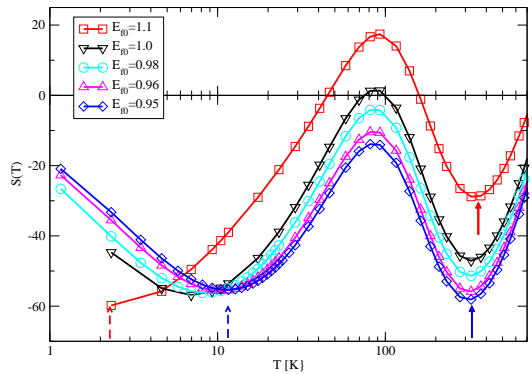}}     
\caption{
The temperature dependence of $S(T)$ is shown for several values of $E_f^0$, as indicated in the figure. 
The other parameters are the same as in Fig.\ref{fig:spectrum_Yb_CF_case}. 
The reduction of $E_f^0$ increases $T_K$, so the low-temperature minimum of $S(T)$ 
(indicated by the dashed arrows)  moves to higher temperatures. 
The effect on the high-temperature minimum (indicated by the arrows) is just the opposite: 
The break-down of the highly degenerate LM regime occurs at slightly lower temperatures. 
 }
                                  \label{fig:thermopower_Yb_various_Ed}      
\end{figure}
\begin{figure}
\center{\includegraphics[width=.8 \columnwidth,clip] {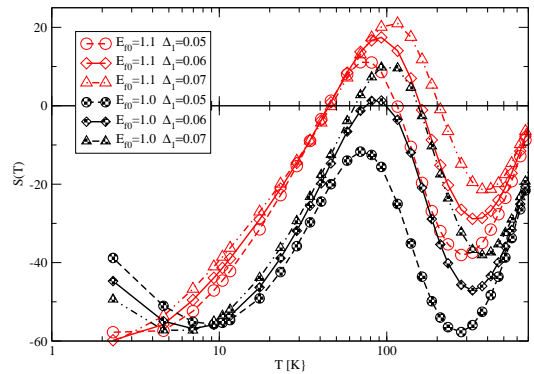}}
\caption{
The dependence of $S(T)$  on the position of the first excited CEF level $\Delta_1$ is shown for 
$E_f^0=1.1$ (red symbols) and $E_f^0=1.0$ (black symbols). 
The data are plotted for $\Delta_1$=0.05 eV (dashed line), $\Delta_1$=0.06 eV (full line), 
and $\Delta_1$=0.07 (dashed-dotted line). 
A simultaneous decrease of $\Delta_1$ and increase of $E_f^0$, due to chemical pressure, 
reduces the Kondo scale and moves $T_0$ to lower temperature, while $T_\mathrm{min}$ doesn't change much. 
 }
                                  \label{fig:thermopower_Yb_various_delta}      
\end{figure}

The crossovers shown by the spectral function are also found in the temperature dependence 
of the thermopower plotted in Fig.\ref{fig:thermopower_Yb_various_Ed}  for several values of $E_f^0$. 
(All other parameters  in Fig.\ref{fig:thermopower_Yb_various_Ed} are the same as in Fig.\ref{fig:spectrum_Yb_CF_case}.) 
The modification of $S(T)$ due to a simultaneous change of the exchange coupling and the CEF splitting 
is shown in Fig.\ref{fig:thermopower_Yb_various_delta}. 
The data plotted in Figs.~\ref{fig:thermopower_Yb_various_Ed} and \ref{fig:thermopower_Yb_various_delta} 
reveal the following features. Above 300 K, $S(T)$ is a logarithmic function of temperature with a positive slope
(the logarithmic behavior extends far above 700 K, the highest temperature shown in the figure). 
Below the room temperatures, $S(T)$ has a minimum at $T_\mathrm{min}$, a maximum at $T_\mathrm{max}$, 
and a low-temperature minimum at $T_0$. 
The thermopower at $T_\mathrm{min}$ is more negative for smaller $E_f^0$, i.e., $\vert S(T)\vert$ is large for larger 
coupling (larger Kondo scale). 
As shown in Fig.~\ref{fig:thermopower_Yb_various_Ed}, 
for a given CEF splitting, a reduction of $E_f^0$ increases $T_0$ and reduces $T_\mathrm{min}$, 
so that the separation between the two minima decreases and the double-well shape of $S(T)$ is less pronounced.  
When  $E_f^0$ is small enough, i.e.,  the 4$f$-state is sufficiently close to the chemical potential, 
the thermopower minima coalesce and, eventually, a single broad minimum emerges (see Ref.[\onlinecite{zlatic.05}]). 
The thermopower of this shape is typical of  valence fluctuators.

The scaling solution provides a simple physical meaning to the seemingly complicated behavior of $S(T)$ 
shown in Figs.~\ref{fig:thermopower_Yb_various_Ed} and  \ref{fig:thermopower_Yb_various_delta}. 
Since we have $T_K\ll T_K^{\cal N}\ll T_\Delta$, the system has  more than one scaling regime and the thermopower 
has more than one minimum. 
At very high temperatures, the excited CEF states are thermally occupied and the system is fully degenerate.  
Here, the CEF splitting  enters only through the renormalization of the Kondo scale $T_K^{\cal N}$ 
and $S(T)$ is a logarithmic function of reduced  temperature $T/T_K^{\cal N}$. 
For $T <T_\Delta$, the excited CEF states depopulate and the effective degeneracy of the $f$-level  decreases. 
The thermopower minimum at $T_\mathrm{min} \simeq T_\Delta\gg T_K^{\cal N}$ indicates the collapse of 
the highly degenerate LM regime and the crossover to the low-temperature LM regime. 
The shift of $T_\mathrm{min}$ with $E_f^0$, which can be seen in Fig.\ref{fig:thermopower_Yb_various_Ed}, 
is also explained by scaling. For a given CEF splitting, an increase of the exchange coupling 
(reduction of $E_f^0$) enhances the Kondo scale, so the quantum fluctuations diminish 
the crystal field effects and shift the ${\cal N}$-fold degenerate regime to lower temperature. 

In the crossover region, $T<T_\mathrm{min}$, the thermopower increases until the low-temperature LM state 
is stabiilized, as indicated by the maximum of $S(T)$ at $T_\mathrm{max}$. 
The value of $S(T)$ around $T_\mathrm{max}$ is negative for small $E_f^0$  (large coupling) and positive 
for large enough $E_f^0$  (small coupling). 
Below $T_\mathrm{max}$, the thermopower behaves as expected of a twofold degenerate $f$ hole: 
it decreases logarithmically toward the low-temperature minimum at $T_0$. 
This minimum signifies the collapse of the twofold degenerate LM state and the crossover into the FL regime;  
it is visible in Fig.\ref{fig:thermopower_Yb_various_Ed} for $E_f^0\leq 1.0$, while 
for $E_f^0 > 1.0$, our numerical procedure becomes unstable before we reach $T_0$. 
At lowest temperatures, such that $T\ll T_K$, the thermopower should follow the FL law, 
$S(T)\simeq - T/T_K$. However, we cannot reach that regime by our NCA solution.  
For weak coupling, the temperature at which $S(T)$ changes sign provides an estimate of $T_K$ 
which agrees with the Kondo scale inferred from the maximum of the spectral function. 
For larger coupling, the low-temperature thermopower doesn't change sign and 
we obtain an order of magnitude estimate for the Kondo scale of the effective doublet 
by using of $T_K\propto T_0$. 

If  $E_f^0$ is reduced sufficiently, the exchange coupling becomes such that $T_K^{\cal N}\geq T_\Delta$, 
and quantum fluctuations render the CEF splitting irrelevant,  i.e., the system behaves as effectively 
${\cal N}$-fold degenerate regardless of temperature. 
In that case, the NCA calculations\cite{Bickers1987,zlatic.05} yield the thermopower with 
a single negative minimum, the shape of which depends on $\cal N$.
For  large degeneracy, ${\cal N}\gg 1$, the thermopower is always negative and its high-temperature slope 
is very small; the minimum of $S(T)$ is barely visible and the value of $T_\mathrm{min}$ is 
unrelated to $T_K^{\cal N}$. When degeneracy is reduced, ${\cal N} \leq 6$,  the thermopower changes sign 
at high enough temperatures and the minimum of $S(T)$ is better pronounced. 
For $\cal N$=2, the temperature at which $S(T)$ changes sign or the temperature of the minimum of $S(T)$  
can be simply related to the Kondo scale of the model.

The variation of $S(T)$ due to the change of the CEF splitting is presented in  Fig.\ref{fig:thermopower_Yb_various_delta}. 
The data show that a  decrease of $\Delta_1$, with other parameters being fixed, shifts $T_\mathrm{min}$ to lower 
and $T_0$ to higher values. This is consistent with the scaling analysis which shows that the reduction of  
the CEF splitting has a two-fold effect: (i) it reduces temperature at which the excited CEF states  start 
to depopulate, thereby decreasing $T_\mathrm{min}$, (ii) it enhances the Kondo scale of the effective doublet, 
so $T_0$ increases. (For an effective doublet, $T_0$ is proportional to $T_K$.)   
On the other hand, an increase of  $E_f^0$, with other parameters being fixed, has the opposite effect. 
It reduces  the exchange coupling and shifts $T_K$ and $T_0$ to lower values. 
But the effect of quantum fluctuations is now also reduced, so higher temperature is needed 
to populate the excited CEF states, i.e.,  $T_\mathrm{min}$ increases. 

If an application of chemical pressure reduces simultaneously the CEF splitting and the exchange coupling 
(by increasing $E_f^0$), the position of $T_\mathrm{min}$ doesn't change much, while $T_0$ 
moves to lower temperatures. [Compare the data obtained for $E_f^0$=1.0, $\Delta_1$=0.06 
(black diamonds) and $E_f^0$=1.1, $\Delta_1$=0.05 (red circles), showing that in both cases 
$T_\mathrm{min}$ is nearly the same, while the shift of $T_0$ is large.]

\section{Comparison of the theory and experiment
\label{comparioson}  }

The single impurity Anderson model presented in the previous sections captures 
the essential features of the thermopower of the Yb(Rh$_{1-x}$Co$_x$)$_2$Si$_2$ series of compounds, 
if we assume that the substitution of rhodium by cobalt increases chemical pressure. 
In Yb-based compounds, this  has a two-fold effect. First, it shifts the 4$f$ level  away from the chemical 
potential, so as to reduce the exchange coupling and the Kondo scale. 
Second, it  decreases the CEF splitting, thereby  lowering the temperature at which thermal fluctuations  
change the occupation of the excited CEF states and the effective degeneracy of 4$f$ states. 
Since the Kondo scale depends exponentially on the exchange coupling,  
while the CEF splitting is assumed to be linear in cobalt concentration, 
their ratio is a highly non-linear function of cobalt concentration.
As a consequence, an increase of cobalt  has a large impact on the properties of the system. 
It brings us from the case where $T_K/\Delta_1\simeq 1$ and $S(T)$ has a single minimum 
to the case where $T_K/\Delta_1\ll 1$ and $S(T)$ acquires a secondary minimum due to the 
thermal occupation of excited CEF states. This behavior is discussed in more detail below. 

At high cobalt concentration, $1\geq x \geq 0.68$, the experimental data are described by the model 
with the input parameters such that $T_K\ll  T_K^{\cal N}\ll \Delta_1$.  
For $x=1$, we have the highest value of $E_f^0$ (the lowest coupling) and the lowest value of $\Delta_1$. 
In this weak-coupling regime, the thermopower exhibits a negative minimum at temperature 
$T_0<10$ K. This minimum indicates a crossover between a non-degenerate FL ground state and 
a twofold degenerate LM regime with Kondo scale $T_K\propto T_0$. 
The reduction of cobalt concentration increases $T_0$, in agreement with the NCA results 
which yield a larger $T_K$ for smaller $E_f^0$. 
Above the minimum, $T>T_0$, the thermopower becomes less negative and,  
at high enough temperature,  $S(T)$  attains a local maximum, where $S(T)>0$ for $x\simeq 1$ 
and $S(T)<0$ for $x<1$. 
This maximum signifies the collapse of the twofold degenerate LM regime 
and the onset of a broad crossover to the high-temperature regime, where the excited CEF states 
are occupied due to thermal fluctuations and the 4$f$ state is $\cal N$-fold degenerate. 
The higher the cobalt concentration, the lower the values of $T_0$ and temperature at which $S(T)$ 
changes sign. The effect of the CEF splitting on these temperatures is completely accounted for by the change 
it induces on $T_K$, as inferred from the maximum of the spectral function. 

The crossover region terminates at temperature $T_\mathrm{min} $, where $S(T)$ has a deep (negative) minimum. 
For $T > T_\mathrm{min}$, the $f$-state is effectively $\cal N$-fold degenerate and the Kondo scale is $T_K^{\cal N}\gg T_K$. 
Here, the slope of the thermopower is positive, so that $S(T)$ can change sign at high enough temperature. 
Unlike the minimum at $T_0$, which is related to $T_K$, the minimum at $T_\mathrm{min}$ is caused 
by an interplay of quantum and thermal fluctuations and it cannot be simply related to $T_K^{\cal N}$. 
A simultaneous increase of $T_K^{\cal N}$ and $\Delta_1$, due to the reduction of cobalt concentration, 
doesn't necessarily change $T_\mathrm{min}$, because an enhancement of quantum fluctuations 
(larger exchange coupling), and diminished effect of thermal fluctuations (the excited CEF states being 
less accessible), shift $T_\mathrm{min}$ in opposite directions. 
This feature is found in the experimental data (see Fig.\ref{fig:thermopower})  which show 
that a change in cobalt concentration changes the width of the crossover region, $T_\mathrm{min}-T_\mathrm{max}$, 
by shifting $T_\mathrm{max}$ but not $T_\mathrm{min}$. 
In model calculations (see Fig.\ref{fig:thermopower_Yb_various_delta}), the comparison of the data set 
($E_f^0=1.1, \Delta_1=0.05$), red circles, and the data set  ($E_f^0=1.0, \Delta_1=0.06$), black diamonds, 
shows that $T_0$ and $T_\mathrm{max}$ increase, while $T_\mathrm{min}$ doesn't change as we  increase $\Delta_1$ 
and reduce $E_f^0$. 

At intermediate concentrations, $0.68>x>0.195$, the experiments are explained by the model in which 
the exchange coupling and the CEF splitting are further increased with respect to the $x\simeq 1$ case. 
This makes the thermopower at $T_\mathrm{max}$ and $T_\mathrm{min}$ more negative but it also brings $T_\mathrm{max}$ 
closer to $T_\mathrm{min}$, so that the double-well shape of $S(T)$ becomes less pronounced. 
In the intermediate coupling regime, the parameters are such that 
$T_K <  T_K^{\cal N} \leq T_\Delta$.  
For low enough $x$ (smaller $E_f^0$ and larger $\Delta_1$), the maximum of $S(T)$ is suppressed 
completely, the value of $S(T)$ at $T_\mathrm{min}$ is large and negative, and the double-well shape of $S(T)$ 
is transformed into a single minimum with a shoulder on the low-temperature side\cite{zlatic.05}.
 
At low cobalt concentration, $x < 0.195$, we assume that the coupling constant is so large that 
quantum fluctuations dominate over thermal fluctuations and the CEF splitting doesn't play any role.  
Regardless of temperature, the $f$-state behaves as an effective octet with Kondo scale 
$T_K^{\cal N}$ which is comparable to $\Delta_1$.
For $T\geq T_K^{\cal N}$, the system has a huge magnetic entropy and large (negative) 
thermopower with positive slope. To remove the magnetic  entropy at low temperatures, 
for $T\ll T_K^{\cal N}$, the system makes a crossover from an $\cal N$-fold degenerate LM state 
to a non-degenerate FL ground state, where the thermopower follows the FL law, $S(T)\propto -T$.
Thus, the crossover gives rise to the thermopower with a single minimum.  
An increase of the exchange coupling, at constant $\cal N$, enhances the Kondo scale and 
broadens the minimum of $S(T)$; it also reduces the high-temperature slope of $S(T)$, 
so as to make the minimum of $S(T)$  less pronounced. 
For large $\cal N$, the values of $T_\mathrm{min}$ and $T_K^{\cal N}$ are not simply related.
The thermopower can changes sign at very high temperatures, except at lowest cobalt concentrations, 
where $S(T)$ is always  negative.  Behaviour of this type is seen not just in  
Yb(Rh$_{1-x}$Co$_x$)$_2$Si$_2$ for low cobalt concentration but also in other Yb systems 
with large Kondo scale\cite{TB-99-3,YRS-08-4,TB-12-2,TB-99-4}. 

In cerium compounds, where pressure (or chemical pressure) increases the exchange coupling, 
the overall behavior of transport coefficients is also explained by 
an interplay of the Kondo effect and CEF splitting.\cite{zlatic.2014} Here, the 4$f$ ions fluctuate 
between  the 4$f^0$ and 4$f^1$ configurations and the thermopower  looks like a mirror image of what one finds in Yb compounds. 
For example\cite{wilhelm.2005}, the thermopower of CeRu$_2$Ge$_2$ exhibits, at ambient pressure, two well resolved maxima: 
the low-temperature  one, indicating the Kondo effect due to an effective doublet, and the high-temperature one, indicating 
a crossover to a fourfold or sixfold degenerate $f$-state. 
An application of pressure enhances the hybridization and increases the Kondo scale 
but doesn't change temperature at which the CEF states become uniformly occupied.
Thus, the low-temperature (Kondo) maximum of $S(T)$ moves to higher temperatures and the crossover 
region shrinks. In CeRu$_2$Ge$_2$, the CEF splitting becomes irrelevant above 8 GPa, regardless of temperature,  
and $S(T)$ exhibits a single broad maximum. Similar behavior is found in the Ce$_x$La$_{1-x}$Cu$_2$Si$_2$ series 
of compounds\cite{ocko.2001}, where Ce substitution reduces the lattice constant. 
The ensuing chemical pressure enhances the exchange coupling, so the two thermopower maxima,  
observed at the lanthanum-rich end of the series, become less pronounced. 

\section{Conclusions  and summary
\label{conclusions}  }
The dependence of transport coefficients of the Yb(Rh$_{1-x}$Co$_x$)$_2$Si$_2$ series of alloys 
on temperature and cobalt concentration is explained by an asymmetric Anderson model 
which  takes into account the exchange scattering of conduction electrons on ytterbium ions 
and the splitting of the 4$f$-state by the CEF into several multiplets. 
The substitution of rhodium  by cobalt is treated as an increase of chemical pressure which 
reduces the exchange coupling and the CEF splitting. 
The model neglects all the details of the band structure, so it cannot provide a quantitative 
description of the experimental results. For such a description, one would also need the single 
ion contribution to measured thermopower data, which is difficult to get because, in Kondo systems,  
the  Nordheim-Gorter rule doesn't hold. However, on a qualitative level, the model captures 
all the main features of the experiment.

The scaling analysis and the numerical solutions of NCA equations show that the effective 
degeneracy of the 4$f$-state depends, at a given temperature, on the relative importance 
of quantum fluctuations, driven by the exchange scattering, and thermal fluctuations, populating 
the excited CEF states. The properties of such a system crucially depend on the ratio $T_K/\Delta$. 
Since the CEF splitting and the exchange coupling are linear in cobalt concentration, while 
$T_K$ is an exponential function of the exchange coupling, the ratio $T_K/\Delta$ is a highly 
non-linear function of cobalt concentration. 
Plotting $T_K$ and $\Delta$ versus chemical pressure yields, in analogy with the Doniach diagram,
a phase diagram in which the weak coupling and the strong coupling regions can be distinguished.  
The temperature dependence of the response functions in these two regions is completely different: 
In the strong coupling case, the system makes a single crossover from a high-temperature, 
$\cal N$-fold degenerate LM state to a non-degenerate FL ground state and the thermopower 
exhibits a single minimum.  In the weak coupling case,  there are two crossovers: first, from 
the $\cal N$-fold to the twofold degenerate LM state and, then, to the FL ground state. 
Here, the thermopower exhibits two minima. 
Thus, the evolution of thermopower across the Yb(Rh$_{1-x}$Co$_x$)$_2$Si$_2$ series 
of compounds provides an insight in the interplay of quantum and thermal fluctuations which 
depends on the relative magnitude of the Kondo scale and the CEF splitting.
Similar analysis holds for many other intermetallic compounds with ytterbium and cerium ions. 

\acknowledgments 
V.Z. acknowledges support by the Ministry of Science of Croatia under the bilateral 
agreement with the USA on scientific and technological cooperation, Project No. 1/2016.
\bibliography{bibliography_file.bib}
\end{document}